\documentstyle[multicol,aps,prl,floats,ifthen]{revtex}
\tighten

\newcommand{\wide}[2]{                                                        %
\end{multicols}                                                               %
\widetext                                                                     %
\noindent                                                                     %
\ifthenelse{\equal{#1}{t}}                                                    %
{}                                                                            %
{                                                                             %
\raisebox{0.1in}[0in][0.02in]{$\rule{3.575in}{0.002in}                        %
\rule{0.002in}{0.08in}$}                                                      %
}                                                                             %
#2                                                                            %
\ifthenelse{\equal{#1}{b}}                                                    %
{}                                                                            %
{                                                                             %
{\raisebox{-0.1in}[0in][0.02in]                                               %
{\hspace{3.35in}$\rule{0.002in}{0.08in}                                      %
\rule[0.08in]{3.575in}{0.002in}$}                                             %
}                                                                             %
}                                                                             %
\begin{multicols}{2}                                                          %
\noindent                                                                     %
}                                                                             %

\begin{document}

\title{Long-range magnetic interaction due to the Casimir effect}

\author{P. Bruno}

\address{Max-Planck-Institut f\"ur Mikrostrukturphysik,
Weinberg 2, D-06120 Halle, Germany}


\maketitle

\begin{abstract}
The zero-point quantum fluctuations of the electromagnetic field
in vacuum are known to give rise to a long-range attractive force
between metal plates (Casimir effect). For ferromagnetic layers
separated by vacuum, it is shown that the interplay of the Casimir
effect and of the magneto-optical Kerr effect gives rise to a
long-range magnetic interaction. The Casimir magnetic force is
found to decay as $D^{-1}$ in the limit of short distances, and as
$D^{-5}$ in the limit of long distances. Explicit expressions for
realistic systems are given in the large and small distance
limits. An experimental test of the Casimir magnetic interaction
is proposed.\\
\\
published in: Phys. Rev. Lett. {\bf 88}, 240401 (2002)
\end{abstract}

\begin{multicols}{2}
Since the discovery of magnets by the ancient Greeks, long-range
magnetic interactions have been an object of fascination. It is
usually considered that there exists essentially two kinds of
magnetic interactions between magnetic moments or magnetized
bodies: (i) the dipole-dipole magnetostatic interaction, and (ii)
the electron-mediated exchange interaction. The latter have
recently received a renewed attention, with the discovery of a
spectacular oscillatory behavior of the interlayer exchange
coupling between ferromagnetic layers separated by a non-magnetic
metal spacer \cite{ Parkin1990}, due to a spin-dependent quantum
size effect \cite{ Bruno1993}.

For the case of two uniformly magnetized ferromagnetic plates (of
infinite lateral extension) held parallel to each other in
vacuum, the two above mentioned magnetic coupling mechanisms
yield a magnetic interaction which decreases exponentially with
interplate distance $D$: (i) the stray field due to a uniformly
magnetized plate decreases exponentially (with a characteristic
decay length of the oder of the interatomic distance) with the
distance from the plate, and so does also the interplate dipolar
interaction; (ii) the interplate exchange interaction also decays
exponentially with $D$, since it is mediated by electrons
tunneling through vacuum between the two plates. The aim of the
present Letter is to point out the existence of a novel, so far
overlooked, mechanism of magnetic interaction between magnetized
(i.e., ferro- or ferrimagnetic) bodies. This interaction arises
from the Casimir effect, and gives rise to a long-range (i.e.,
with power-law decay) magnetic interaction; at sufficiently large
distance $D$ it is therefore the dominant mechanism of magnetic
interaction between the two ferromagnetic plates.

As Casimir pointed out in a seminal paper \cite{ Casimir1948},
the zero-point quantum fluctuations of the electromagnetic (EM)
field in vacuum leads to observable effects: changing the boundary
condition of the EM field (e.g., by moving a body with respect to
another) yields a finite change of the (infinite) zero-point
energy of the system, and therefore results in an observable
force. In particular, Casimir predicted the existence of a
long-range attractive force between mirrors in vacuum. The Casimir
effect is currently attracting considerable interest \cite{
Chan2001, Spruch1996}, in particular with respect to
micromechanical devices, and has deep implications in many fields
of physics.

The boundary condition of the EM field can, however, also be
modified without any mechanical displacement, but rather, by
changing the order parameter of a collective ordering phenomenon
such as ferromagnetism. When the two mirrors are ferromagnetic,
the magneto-optical Kerr effect influences the boundary condition
of the EM field, so that the Casimir effect manifests as a energy
difference (per unit area) $\Delta {\cal E} \equiv {\cal E}_{AF}
- {\cal E}_{FM}$ between the configurations in which the two
mirrors have their magnetizations antiparallel (AF) or parallel
(FM) to each other, i.e., as a magnetic interaction.
Alternatively, the above effect manifests as a dependence of the
(mechanical) Casimir force (per unit area) among the mirrors upon
the relative orientation of their magnetizations: $\Delta {\cal F}
\equiv {\cal F}_{AF} - {\cal F}_{FM} = -{\rm d}\Delta {\cal E} /
{\rm d}D$. The {\em Casimir magnetic force} $\Delta {\cal F}$ has
been calculated for ferromagnetic mirrors described by a Drude
model: in the limit of very large distance, the magnetic force
decays as $D^{-5}$; in the limit of small distances, it decays as
$D^{-1}$; in the intermediate regime, it decays as $D^{-4}$. For
equivalent ferromagnetic materials on both sides, the Casimir
magnetic interaction is always antiferromagnetic. For realistic
systems, the explicit expression of the Casimir magnetic force is
found to be
%
\begin{equation}\label{eq_force_long}
\Delta {\cal F} \approx  \frac{-3\zeta(3)}{16 \pi^3} \,
\frac{\hbar c^2}{D^5} \, \frac{\theta_A \theta_B}{\sqrt{\sigma_A
\sigma_B}} ,
\end{equation}
in the limit of large distance, where $\sigma_{A(B)}$ and
$\theta_{A(B)}$ are, respectively, the dc conductivity and
anomalous Hall angle of ferromagnetic plate $A$ (resp. $B$), and
\begin{equation}\label{eq_force_short}
\Delta {\cal F} \approx  \frac{-1}{4\pi^2} \, \frac{\hbar}{c^2 D}
\int_0^{+\infty}\!\!\!\!\!\!\!\!\!  \frac{\omega^2 \,
\varepsilon_{xy}^A ({\rm i}\omega ) \, \varepsilon_{xy}^B ({\rm
i}\omega )}{ \left[ 1 + \varepsilon_{xx}^A({\rm i}\omega
)\right]\left[ 1 + \varepsilon_{xx}^B({\rm i}\omega )\right] }\,
{\rm d}\omega ,
\end{equation}
in the limit of short distances, where
$\varepsilon_{xx}^{A(B)}({\rm i}\omega )$ and
$\varepsilon_{xy}^{A(B)} ({\rm i}\omega )$ are, respectively, the
diagonal and off-diagonal elements of the dielectric tensor of
ferromagnetic plate $A$ (resp. $B$), evaluated at imaginary
frequency ${\rm i}\omega$.

The Casimir interaction energy (per unit area) between two
mirrors can be conveniently expressed in terms of their
reflection coefficients as \cite{ Katz1977, Jaekel1991},
\wide{t}{
\begin{equation}\label{eq_energy1}
{\cal E}=\int_0^{+\infty} {\rm d}\omega \, \frac{\hbar}{2}
\coth\left( \frac{\hbar\omega}{k_BT}\right) \, \frac{1}{(2\pi)^2}
\int {\rm d}^2 {\bf k}_\| \, \frac{1}{\pi} \mbox{Im Tr} \ln\left(
1- {\sf R}_A {\sf R}_B {\rm e}^{2ik_\bot D} \right) ,
\end{equation}
}  
where ${\bf k}_\|$ and $k_\bot$ are the components of the
wavevector, respectively, parallel and perpendicular to the
mirrors. The above expression is completely analogous to the one
derived independently for the case of interactions mediated by
fermions (electrons) \cite{ Bruno1993}. The $2\times 2$ matrix of
reflection coefficients on mirror $A$ (resp. $B$) is given by
\begin{equation}
{\sf R}_{A(B)} \equiv \left(
\begin{array}{cc}
r_{ss}^{A(B)} \ \ r_{sp}^{A(B)} \\
r_{ps}^{A(B)} \ \ r_{pp}^{A(B)}  \\
\end{array}
\right) ,
\end{equation}
where the index $s$ (resp. $p$) corresponds to a polarization
with the electric field perpendicular (resp. parallel) to the
incidence plane. The off-diagonal matrix elements $r_{sp}$ and
$r_{ps}$ are responsible for the magneto-optical effects. With the
usual convention that the $s$ axis remains unchanged upon
reflection, one has $r_{sp}=r_{ps}$, and, for perpendicular
incidence, $r_{ss} =- r_{pp}$. Performing the change of variables
$(\omega , {\bf k}_\| ) \to (\omega , k_\bot )$ and using complex
plane integration methods, one can rewrite Eq.~(\ref{eq_energy1})
at $T=0$ as \cite{ Jaekel1991}
\wide{m}{
\begin{equation}\label{eq_energy2}
{\cal E}=\frac{\hbar}{4\pi^2} \int_0^{+\infty} \!\!\! k_\bot \,
{\rm d}k_\bot \int_0^{k_\bot c} \! \! {\rm d}\omega \, \mbox{Re
Tr } \ln\left[ 1- {\sf R}_A({\rm i}\omega , {\rm i}k_\bot) \, {\sf
R}_B({\rm i}\omega ,{\rm i}k_\bot) \, {\rm e}^{-2k_\bot D}
\right] ,
\end{equation}
where the reflection coefficients are evaluated at imaginary
values of the frequency and normal wavevector.

For a mirror magnetized along its normal (pointing outwards), the
reflection coefficients are given by \cite{ Zak1990}
\begin{mathletters}
\begin{eqnarray}
\label{eq_refl1}
r_{ss}({\rm i}\omega ,{\rm i}k_\bot) &=&
\frac{k_\bot c - \sqrt{ \omega^2 \left(\varepsilon_{xx} ({\rm
i}\omega ) -1 \right) + \left( k_\bot c \right)^2 } } {k_\bot c +
\sqrt{ \omega^2 \left(\varepsilon_{xx} ({\rm i}\omega ) -1
\right) + \left( k_\bot c \right)^2 } }, \ \ \ r_{pp}({\rm
i}\omega ,{\rm i}k_\bot) = \frac{\varepsilon_{xx} ({\rm i}\omega
) k_\bot c - \sqrt{ \omega^2 \left(\varepsilon_{xx} ({\rm
i}\omega ) -1 \right) + \left( k_\bot c \right)^2 } }
{\varepsilon_{xx} ({\rm i}\omega ) k_\bot c + \sqrt{ \omega^2
\left(\varepsilon_{xx} ({\rm i}\omega ) -1 \right) + \left(
k_\bot c \right)^2 } } \\
\label{eq_refl2}
r_{sp}({\rm i}\omega ,{\rm i}k_\bot) &=&
r_{ps}({\rm i}\omega ,{\rm i}k_\bot) = \frac{-k_\bot c\, \omega\,
\varepsilon_{xy} ({\rm i}\omega )}{ \left[ k_\bot c + \sqrt{
\omega^2 \left(\varepsilon_{xx} ({\rm i}\omega ) -1 \right) +
\left( k_\bot c \right)^2 }\right] \left[ \varepsilon_{xx} ({\rm
i}\omega ) k_\bot c + \sqrt{ \omega^2 \left(\varepsilon_{xx}
({\rm i}\omega ) -1 \right) + \left( k_\bot c \right)^2 }\right]
} .
\end{eqnarray}
\end{mathletters}
For the sake of simplicity, the arguments $({\rm i}\omega ,{\rm
i}k_\bot)$ will be omitted below. If the magnetization point
inwards, then the sign of $r_{sp}$ and $r_{ps}$ is reversed.

As the magneto-optical reflection coefficients $r_{sp}$ are
usually much smaller than 1 and than the usual reflection
coefficients $r_{ss}$ and $r_{pp}$, one can expand the Casimir
magnetic energy to lowest order in the magneto-optical
coefficients, yielding
\begin{equation}\label{eq_magn_energy1}
\Delta {\cal E} = \frac{-\hbar}{\pi^2} \int_0^{+\infty} \!\!\!
k_\bot \, {\rm d}k_\bot \int_0^{k_\bot c} \! \! {\rm d}\omega \,
\mbox{ Re} \left[ \frac{ r_{sp}^A r_{sp}^B {\rm e}^{-2k_\bot D}}
{\left( 1 -r_{ss}^A r_{ss}^B {\rm e}^{-2k_\bot D} \right) \left( 1
-r_{pp}^A r_{pp}^B {\rm e}^{-2k_\bot D} \right) } \right] .
\end{equation}
}  
Eq.~(\ref{eq_magn_energy1}) together with Eqs.~(\ref{eq_refl1},b)
allow to calculated the Casimir magnetic energy and force.

In order to illustrate the above result, let us calculate the
Casimir magnetic interaction for the case of (equivalent)
ferromagnetic mirrors with a dielectric tensor approximated by a
Drude model:
\begin{equation}
\varepsilon_{xx}({\rm i}\omega) =
1+\frac{{\omega_p}^2\tau}{\omega (1+\omega\tau)} , \ \
\varepsilon_{xy}({\rm i}\omega) = \frac{{\omega_p}^2 \omega_c
\tau^2}{\omega (1+\omega\tau)^2} .
\end{equation}
The plasma frequency $\omega_p$ is given by ${\omega_p}^2 \equiv
4\pi n e^2 / m^\star$; $\omega_c \equiv eB_{\rm eff}/m^\star c$
is the cyclotron frequency, where $B_{\rm eff}$ is the effective
magnetic field experienced by conduction electrons as a result of
the combined effect of the exchange and spin-orbit interactions;
$\tau$ is the relaxation time. It is assumed that $\omega_c \tau
\ll 1 \ll \omega_p\tau$, which constitutes the usual situation.

One can distinguish three different regimes: (i) $D \gg c\tau$,
(ii) $c/\omega_p \ll D \ll c\tau$, (iii) $D \ll c/\omega_p$. In
regime (i) (i.e., at long distances), the integral in
Eq.~(\ref{eq_magn_energy1}) is dominated by the range $\omega
\leq k_\bot c  \approx c/D \ll 1/\tau$, for which one has
\begin{equation}
\varepsilon_{xx}({\rm i}\omega) \approx
\frac{{\omega_p}^2\tau}{\omega} \gg 1 , \ \varepsilon_{xy}({\rm
i}\omega) \approx \frac{{\omega_p}^2\omega_c\tau^2}{\omega} ,
\end{equation}
so that
\begin{equation}\label{eq_refl_long}
r_{ss} \approx -r_{pp} \approx -1 , \ \ r_{sp} \approx
-\frac{\omega_c}{\omega_p}\sqrt{\omega\tau} .
\end{equation}
One then finds that
\begin{equation}
\Delta {\cal E} \approx \frac{-3\zeta(3)}{16\pi^2} \, \frac{\hbar
c^2}{D^4}  \, \frac{{\omega_c}^2\tau}{{\omega_p}^2} , \ \Delta
{\cal F} \approx \frac{-3\zeta(3)}{4\pi^2} \, \frac{\hbar
c^2}{D^5}  \, \frac{{\omega_c}^2\tau}{{\omega_p}^2} ,
\end{equation}
where $\zeta(x) \equiv \sum_{n=1}^\infty n^{-x}$ is the Riemann
zeta function, with $\zeta(3) \approx 1.202\dots$

In the intermediate distance regime (ii) ($c/\omega_p \ll D \ll
c\tau$), the integral in Eq.~(\ref{eq_magn_energy1}) is dominated
by the range $1/\tau \ll\omega \leq k_\bot c  \approx c/D \ll
\omega_p$, for which one has
\begin{equation}\label{eq_eps_xy_interm}
\varepsilon_{xx}({\rm i}\omega) \approx
\frac{{\omega_p}^2}{\omega^2} \gg 1 , \ \
 \varepsilon_{xy}({\rm i}\omega) \approx
\frac{{\omega_p}^2\omega_c}{\omega^3} ,
\end{equation}
so that
\begin{equation}\label{eq_refl_interm}
r_{ss} \approx -r_{pp} \approx -1 , \ \ r_{sp} \approx
-\frac{\omega_c}{\omega_p} .
\end{equation}
One then finds that
\begin{equation}
\Delta {\cal E} \approx -\, \frac{1}{24} \, \frac{\hbar c}{D^3}
\, \frac{{\omega_c}^2}{{\omega_p}^2} , \ \ \Delta {\cal F}
\approx -\, \frac{1}{8} \, \frac{\hbar c}{D^4} \,
\frac{{\omega_c}^2}{{\omega_p}^2} .
\end{equation}

In the short distance regime (iii) ($D \ll c/\omega_p$), one needs
to consider separately the range with $\omega\leq k_\bot c \ll
\omega_p$, for which the reflection coefficients are given by
Eqs.~(\ref{eq_refl_interm}), and the range with $\omega_p \ll
\omega \leq k_\bot c$. For $\omega\gg\omega_p$,
$\varepsilon_{xy}$ is given by Eq.~(\ref{eq_eps_xy_interm}) and
\begin{equation}
\varepsilon_{xx}({\rm i}\omega) -1 \approx
\frac{{\omega_p}^2}{\omega^2} \ll 1 ,
\end{equation}
so that, for $k_\bot c\gg \omega_p$,
\begin{equation}\label{eq_refl_short}
|r_{ss}| \ll  1 , \ \ r_{pp} \ll  1 , \ \ r_{sp} \approx
-\frac{{\omega_p}^2\omega_c}{2k_\bot c (2\omega^2 + {\omega_p}^2
)} .
\end{equation}
One then finds that
\begin{mathletters}
\begin{eqnarray}
\Delta {\cal E} &\approx& -\, \frac{1}{16\pi\sqrt{2}}\,
\frac{\hbar}{c^2}\, \ln\left(\frac{c}{\omega^\star D}\right) \,
{\omega_c}^2 \omega_p , \\
\Delta {\cal F} &\approx& -\, \frac{1}{16\pi\sqrt{2}} \,
\frac{\hbar}{c^2D} \, {\omega_c}^2 \omega_p ,
\end{eqnarray}
\end{mathletters}
where $\omega^\star$ is a cut-off frequency of the order of the
plasma frequency $\omega_p$.

For realistic systems, it is in general necessary to perform a
detailed calculation. However, in the limit of large and small
distances, explicit expressions can be obtained. At large
distances (i.e., for $D\gg c/\tau$), the Casimir magnetic force is
essentially determined by the dielectric tensor
$\varepsilon_{ij}({\rm i}\omega)= \delta_{ij}+4\pi
\sigma_{ij}({\rm i}\omega)/\omega$ at low imaginary frequency. In
this regime, one can safely approximate the conductivity tensor
$\sigma_{ij}({\rm i}\omega)$, in the above expression, by its dc
value: $\sigma_{xx}({\rm i}\omega )\approx \sigma$,
$\sigma_{xy}({\rm i}\omega)\approx\sigma\theta$, where $\sigma$
is the dc conductivity, and $\theta$ the anomalous Hall angle of
the ferromagnetic mirror. Proceeding as for the Drude model, one
then obtains,
\begin{equation}
\Delta {\cal E} \approx \frac{-3\zeta(3)}{64 \pi^3} \,
\frac{\hbar c^2}{D^4} \, \frac{\theta_A \theta_B}{\sqrt{\sigma_A
\sigma_B}} ,
\end{equation}
from which Eq.~(\ref{eq_force_long}) follows immediately.

For short distances, the Casimir magnetic interaction is
dominated by imaginary wavevectors ${\rm i}k_\bot$ with
$\omega^\star \ll k_\bot c \approx c /D$, where the cut-off
frequency $\omega^\star$ is of the order of the plasma frequency,
or the typical frequency of interband transitions. In this
regime, one has
\begin{equation}
|r_{ss}| \ll  1 , \ \ r_{pp} \ll  1 , \ \ r_{sp} \approx  -\,
\frac{\omega \, \varepsilon_{xy} ({\rm i}\omega ) }{2k_\bot c
\left[ 1 + \varepsilon_{xx} ({\rm i}\omega )\right]} .
\end{equation}
One eventually obtains,
\begin{equation}
\Delta {\cal E} \approx \frac{-1}{4\pi^2} \frac{\hbar}{c^2} \ln\!
\left( \frac{c}{\omega^\star D}\right) \!\!
\int_0^{+\infty}\!\!\!\!\!\!\!\!\! \frac{\omega^2\,
\varepsilon_{xy}^A ({\rm i}\omega ) \, \varepsilon_{xy}^B ({\rm
i}\omega )}{ \left[ 1 + \varepsilon_{xx}^A({\rm i}\omega
)\right]\left[ 1 + \varepsilon_{xx}^B({\rm i}\omega )\right] }\,
{\rm d}\omega ,
\end{equation}
from which Eq.~(\ref{eq_force_short}) follows immediately.

Let us now discuss whether the novel Casimir magnetic interaction
can be observed experimentally. Obviously, the regime of potential
experimental interest is the short distance limit. To obtain a
rough estimate of the magnitude of the effect, it is sufficient
to approximate the (magneto-)optical absorption spectrum by a
single absorption line at frequency $\omega_0$ containing all the
spectral weight, i.e., we write:
\begin{mathletters}
\begin{eqnarray}
\mbox{Im } \varepsilon_{xx} (\omega ) &\approx& \omega_0 \,
\varepsilon_{xx}^{\rm eff} \, \delta (\omega - \omega_0) , \\
\mbox{Re } \varepsilon_{xy} (\omega ) &\approx& \omega_0 \,
\varepsilon_{xy}^{\rm eff} \, \delta (\omega - \omega_0) .
\end{eqnarray}
\end{mathletters}
This is expected to be a good approximation in the limit of small
distances (i.e., high frequencies) where the details of the
(magneto-)optical spectra should not matter too much. The
dielectric tensor at imaginary frequency is then obtained from
the causality relations
\begin{mathletters}
\begin{eqnarray}
\varepsilon_{xx} ({\rm i}\omega ) &=&1+\frac{2}{\pi}
\int_0^{+\infty} \!\!\! {\rm d}\omega' \, \frac{\omega' \, \mbox{
Im }\varepsilon_{xx}(\omega'
)}{{\omega'}^2 + \omega^2} , \\
\varepsilon_{xy} ({\rm i}\omega ) &=& \frac{2}{\omega\pi}
\int_0^{+\infty} \!\!\! {\rm d}\omega' \, \frac{{\omega'}^2 \,
\mbox{ Re }\varepsilon_{xy}(\omega' )}{{\omega'}^2 + \omega^2} ,
\end{eqnarray}
\end{mathletters}
and one eventually obtains
\begin{mathletters}
\begin{eqnarray}
\Delta {\cal E} &\approx&  \frac{-1}{16\pi^3} \, \frac{\hbar}{c^2}
\, \ln \left( \frac{c}{\omega_0 D}\right) \, \frac{{\omega_0}^3
\, {\varepsilon_{xy}^{\rm eff}}^2 }{ \left( 1 +
\varepsilon_{xx}^{\rm
eff} / \pi \right)^{3/2} } , \\
\Delta {\cal F} &\approx&  \frac{-1}{16\pi^3} \, \frac{\hbar}{c^2
D} \, \frac{{\omega_0}^3 \, {\varepsilon_{xy}^{\rm eff}}^2 }{
\left( 1 + \varepsilon_{xx}^{\rm eff} / \pi \right)^{3/2} } .
\end{eqnarray}
\end{mathletters}
By simple inspection of the (magneto-)optical absorption spectra
of transition metal ferromagnets \cite{ Ebert1996}, one finds
that the model parameters assume the typical values $\omega_0
\approx  6 \times 10^{15} \ {\rm s}^{-1}$, $\varepsilon_{xx}^{\rm
eff} \approx  10$, $\varepsilon_{xy}^{\rm eff} \approx  1.5
\times 10^{-2}$. Experimentally, it is usually not convenient to
maintain two plates accurately parallel to each other, so that a
configuration with a planar mirror and a lens-shaped mirror is
usually adopted. Even in this configuration, the parasitic
magnetostatic interaction can be made as small as needed by
taking a uniformly magnetized plate of sufficiently low thickness
and sufficiently large lateral extension. The net resulting
Casimir magnetic force $\Delta F$ (not to be confused with the
Casimir magnetic force {\em per unit area} $\Delta {\cal F}$) is
then obtained by means of the ``proximity force theorem'' \cite{
Blocki1977}: $ \Delta F = 2\pi R \, \Delta {\cal E}(D),$ where
$R$ is the curvature radius of the lens-shaped mirror and $D$ the
shortest distance. For $R=100\ \mu$m and $D\leq c/\omega_0
\approx 50$~nm, one finds that $|\Delta F | \approx 10$~fN, with
only a weak (logarithmic) dependence upon $D$.

The Casimir magnetic force $\Delta F$ is several orders of
magnitude weaker than the non-magnetic Casimir force $F$ ($|F|
\approx 0.5$~nN, for $D=50$~nm); however, it can be detected
independently of $F$ by using a resonant differential method, as
done in ``magnetic resonant force microscopy'' \cite{ Rugar1994,
Sidles1995}: in this approach, one measures the mechanical force
between two magnetic samples, one of them being fixed and
magnetically hard, the other one being magnetically soft and
attached to a high-$Q$ mechanical resonator (of resonance
frequency $\omega_r$) consisting of a micro-cantilever. By
modulating the magnetization of the soft sample by means of an ac
magnetic field (the hard sample remaining unchanged) at $\omega
=\omega_r$, one can detect the magnetic force among the two
samples (i.e., the difference $\Delta F \equiv F_{AF} - F_{FM}$)
with a considerably higher sensitivity than allowed by a dc
measurement of $F_{AF}$ or $F_{FM}$ separately. In addition, when
measuring the non-magnetic Casimir effect, great care has to be
taken to eliminate parasitic electrostatic interactions, which is
done automatically in the approach proposed here. A detailed
discussion of the sensitivity limitations of "magnetic resonant
force microscopy" is given in Ref.~\cite{ Sidles1995}: it is
limited on one hand by the sensitivity in measuring the
deflection of cantilever, which yields $|\Delta F_{\mbox{min}}|
\geq k\, \delta x /Q$ (where $k$ is the cantilever spring
constant, $Q$ the quality factor, and $\delta x$ the deflection
sensitivity), and by the thermomechanical noise on the other
hand, which yields $|\Delta F_{\mbox{min}}| \geq \sqrt{4k_BT\,
\Delta\nu\, k /(\omega_r Q)}$, for a bandwidth $\Delta\nu$. For
the cantilever used in Ref.~\cite{ Rugar1994} ($k=1$~mN/m,
$Q=3000$, $\omega_r=1.4$~kHz), a force sensitivity of 0.5~fN at
room temperature was obtained, to be compared with the
sensitivity of at best 1~pN reported in Ref.~\cite{ Chan2001} for
the dc measurement of the non-magnetic Casimir effect. Prospects
for further improvement in the force sensitivity in "magnetic
resonant force microscopy" up to $\approx 3\times
10^{-17}$~N$/\sqrt{\mbox{Hz}}$ at room temperature and $\approx 4
\times 10^{-18}$~N$/\sqrt{\mbox{Hz}}$ at 4.2~K are discussed in
Ref.~\cite{ Sidles1995}. Indeed, force sensitivity in the
attonewton ($10^{-18}$~N) range has recently been demonstrated
\cite{ Stowe1997}.

The cantilever used in Ref.~\cite{ Rugar1994} consisted of a
50~$\mu$m-long, 5~$\mu$m-wide and 90~nm-thick Si beam terminated
by a square paddle of 30~$\mu$m side-length. Such a cantilever
would be appropriate for the proposed experiment. By depositing a
droplet of polymer on the paddle, it would be possible to produce
a lens-shaped substrate of suitable curvature radius
($\approx$100~$\mu$m), which could then be covered by evaporation
with a thin ($\approx$10~nm) layer of a soft ferromagnet such as
permalloy. Great care has to be taken to minimize the parasitic
magnetostatic interaction: the hard magnetic plate should be as
wide, thin, and magnetically uniform as possible. One should
therefore chose a material with a high coercivity and 100\%\
remanence. A thin ($\approx$~10~nm) layer of CoPt alloy or a
Co/Pt multilayer would be suitable. For a plate radius of 1~cm,
the parasitic magnetostatic force can be estimated to be below 1
attonewton, which is sufficient for the present purpose. Finally,
the parasitic force due to the interaction of the soft ferromagnet
with the ac field would yield a signal at 2$\omega$ and would
therefore be filtered out by lock-in detection.

To conclude, the above discussion suggests that the experimental
test of the novel Casimir magnetic interaction would indeed be
possible by using currently available techniques.

\end{multicols}

\end{document}